\documentclass{iau_FM}
\usepackage{graphicx}

\title[Active Asteroids] 
{Active Asteroids: Main-Belt Comets and Disrupted Asteroids}

\author[Henry H.\ Hsieh]   
{Henry H.\ Hsieh$^{1,2}$
}

\affiliation{$^1$Academia Sinica Institute of Astronomy \& Astrophysics \\ 11F, Astronomy-Mathematics Building, National Taiwan University \\
No.\ 1, Sec.\ 4, Roosevelt Road, Taipei 10617, Taiwan \\ email: {\tt hhsieh@asiaa.sinica.edu.tw} \\[\affilskip]
$^2$Planetary Science Institute \\ 1700 East Fort Lowell Road, Suite 106 \\ Tucson, Arizona 85719, United States of America \\email: {\tt hhsieh@psi.edu}}

\pubyear{2015}
\jname{FM9: Highlights in the Exploration of Small Worlds} 
\editors{D.\ Bockel\'ee-Morvan, ed.}
\begin{document}

\maketitle

\begin{abstract}
The study of active asteroids has attracted a great deal of interest in recent years since the recognition of main-belt comets (which orbit in the main asteroid belt, but exhibit comet-like activity due to the sublimation of volatile ices) as a new class of comets in 2006, and the discovery of the first disrupted asteroids (which, unlike MBCs, exhibit comet-like activity due to a physical disruption such as an impact or rotational destabilization, not sublimation) in 2010.  In this paper, I will briefly discuss key areas of interest in the study of active asteroids.
\keywords{comets: general; minor planets, asteroids; astrobiology; methods: n-body simulations; methods: numerical; techniques: photometric; techniques: spectroscopic; surveys
}
\end{abstract}

\firstsection 
\section{Background}

The first object in the main asteroid belt observed to eject mass like a comet (133P/Elst-Pizarro) was discovered in 1996.  Despite initial debate over whether sublimation or an impact event could best explain 133P's activity, impact scenarios were eventually ruled out by observations of the object's reactivation in 2002, leaving sublimation as the most plausible explanation (Hsieh et al.\ 2004).  Main-belt comets (MBCs), which exhibit comet-like mass loss as the result of the sublimation of volatile ice even though they occupy orbits in the main asteroid belt, and of which 133P was the first known example, were recognized as a new class of comets ten years later in 2006 (Hsieh \& Jewitt, 2006).

Perhaps the most significant shift in this field occurred in 2010 with the discoveries of the first ``disrupted asteroids'', P/2010 A2 and (596) Scheila, whose activity was determined to be due to impact disruption, where a collision by another asteroid produces a visible ejecta cloud, or rotational destabilization, where an asteroid rotates faster than the limit at which self-gravity and internal cohesion can prevent mass loss due to centrifugal forces (e.g., Jewitt et al.\ 2010, 2011; Snodgrass et al.\ 2010; Bodewits et al.\ 2011).
As the term ``main-belt comet'' was initially chosen to refer to objects that are physically ``cometary'' (i.e., containing sublimating ice), the use of a new term, ``active asteroids'' (cf.\ Jewitt 2012), was adopted to refer to all
observationally comet-like objects on asteroid-like orbits,
regardless of the dust ejection mechanisms involved.

Dynamically, active asteroids have orbits with Tisserand parameters with respect to Jupiter of $T_J$$\,>\,$3, where $T_J$$\,>\,$3 for most main-belt asteroids, and $T_J$$\,<\,$3 for classical comets (Kres\'ak, 1972), and have semimajor axes smaller than that of Jupiter, i.e., $a$$\,\leq\,$$a_J$.  While most active asteroids are found in the main asteroid belt, use of $T_J$ as the formal dynamical criterion for identifying active asteroids means that a few near-Earth objects with $T_J$$\,>\,$3 that exhibit comet-like activity are also included.


MBCs have attracted interest due to their potential use as tracers of the ice content of the inner solar system (cf.\ Hsieh 2014).  They may also be useful for investigating hypotheses that objects from the main asteroid belt may have played a significant role in the primordial delivery of water to the terrestrial planets (cf.\ Morbidelli et al.\ 2000, 2012; Mottl et al.\ 2007).  Meanwhile, rotational and impact disruptions of asteroids have been extensively studied using numerical models, laboratory experiments, and large statistical studies (e.g., Holsapple et al.\ 2002; Bottke et al.\ 2005; Walsh et al.\ 2012), but the discoveries of real disrupted asteroids has opened up new opportunities for observationally constraining these models, as well as gaining unique insights into the internal structure and physical properties of these objects (e.g., Hirabayashi et al.\ 2014).

In this proceedings paper, I highlight key areas of interest in this rapidly growing and diverse field of research.  For more detailed discussions of active asteroids, the reader is referred to reviews by Jewitt (2012) and Jewitt et al.\ (2015b).  For more details of MBCs in particular, the reader is referred to Bertini (2011) and Hsieh (2015), the related proceedings paper from IAU Symposium 318, which was also held as part of the XXIXth IAU General Assembly. Discussions of observational diagnostics for distinguishing MBCs from disrupted asteroids can be found in Hsieh et al.\ (2012a) and Jewitt et al.\ (2015b).


\section{Main-Belt Comets}

To date, sublimation-driven activity has actually only been inferred for all objects currently considered to be MBCs.  Despite many attempts using ground-based 8-10~m telescopes and the space-based {\it Herschel} telescope, gas products from sublimation have not yet been successfully spectroscopically confirmed for a MBC.
Most ground-based observations have searched for CN emission at 3889\AA\  (Jewitt et al.\ 2009, 2014, 2015a; Hsieh et al.\ 2012b, 2012c, 2013; Licandro et al.\ 2011, 2013), finding $3\sigma$ upper limit production rates of $Q_{\rm CN}$$\,\sim\,$10$^{21}$$\,-\,$10$^{23}$~mol~s$^{-1}$, from which water production rates of $Q_{\rm H_2O}$$\,<\,$10$^{24}$$\,-\,$10$^{26}$ were inferred, usually assuming JFC-like $Q_{\rm CN}/Q_{\rm H_2O}$ ratios (e.g., A'Hearn et al.\ 1995).
Hypervolatile species like CN may be more depleted than ${\rm H_2O}$ in main-belt objects compared to JFCs though (Prialnik \& Rosenberg 2009).  As such, the real upper limit water production rates implied by these CN limits could be much higher.  {\it Herschel} observations directly targeting the 557 GHz $1_{10}$$\,-\,$$1_{01}$ ground state rotational transition of ${\rm H_2O}$ did set limits of $Q_{\rm H_2O}$$\,<\,$4$\times$10$^{25}$~mol~s$^{-1}$ for 176P/LINEAR and $Q_{\rm H_2O}$$\,<\,$8$\times$10$^{25}$~mol~s$^{-1}$ for P/2012 T1 (de Val-Borro et al.\ 2012; O'Rourke et al.\ 2013). No dust emission was observed for 176P at the time though (Hsieh et al.\ 2014), and P/2012 T1's activity was well past its peak (Hsieh et al.\ 2013), so their water production rates may have both been higher during periods of stronger activity.

Detection of sublimation or even ice on a MBC observed to exhibit visible dust emission is challenging due to the faintness of both MBCs and their associated activity, and the difficulty of scheduling observations at times of peak activity when the chances of detecting sublimation are maximized.  Water vapor has been detected from dwarf planet (1) Ceres (K\"uppers et al.\ 2014) and surface ice has been detected on large main-belt asteroids (24) Themis and (90) Antiope (Rivkin \& Emery 2010; Campins et al.\ 2010; Hargrove et al.\ 2015), but these objects are much larger (diameters of $D>100$~km) and brighter than any of the km-scale MBCs, and visible dust emission has never been observed for any of them.  Meanwhile, the point of peak activity is impossible to immediately determine for newly discovered MBCs, meaning that deep spectroscopic observations of a new MBC can easily be conducted too early or too late. Spectroscopic observations could instead be scheduled for a known MBC during its next perihelion passage, but observability restrictions may apply (e.g., the object could be obscured by the Sun or far from the Earth when activity is expected to peak), the object could exhibit much weaker activity than before, and at least one orbit period must elapse following the first active episode, limiting short-term observing opportunities.  As such, thus far, a successful verification of sublimation on a MBC observed to exhibit visible dust emission remains elusive.

\section{Disrupted Asteroids}

Perhaps the most unambiguous example of an impact-driven dust ejection event is that observed for Scheila in 2010, where numerical dust modeling of radiation pressure acting on a hollow ejecta cone and a down-range plume from an oblique impact by a decameter-sized impactor provided an excellent match to the observed multi-plumed dust structure around the asteroid (Ishiguro et al.\ 2011).  Dust emission events for P/2010 A2 and P/2012 F5 (Gibbs) were both initially assumed to be caused by impacts (e.g., Jewitt et al.\ 2010; Snodgrass et al.\ 2010; Stevenson et al.\ 2012), but later analyses suggest that both events may have been caused by rotational destabilization instead (Agarwal et al.\ 2013; Drahus et al.\ 2015).  Other objects suspected of being rotationally disrupted have also been identified (e.g. 311P/PANSTARRS, P/2013 R3, and (62412) 2000 SY$_{178}$; Jewitt et al.\ 2013, 2014; Sheppard \& Trujillo 2015), although of these, only (62412) has had its rotation period determined and been confirmed to be a fast-rotator.

Disruption events provide opportunities to probe the physical properties of asteroids in various ways.  Impact events like the one experienced by Scheila may alter the surface of the impacted asteroid, depending on the size of the impactor and other details of the impact event, where these changes can potentially then be used to infer properties of the impacted surface material by placing observational constraints of physical models of the impact, as well as providing direct compositional information about the excavated material (e.g., Bodewits et al.\ 2014).  In principle, the rates at which impacts occur could also place constraints on the otherwise unobservable population of m- or 10-m-scale main-belt asteroids (cf.\ Hsieh 2009).  By placing constraints on numerical models and laboratory experiments, impact and rotational disruption events can also be used to infer material properties such as the internal strength and porosity of the affected asteroid (e.g., Housen \& Holsapple 2003; Holsapple 2009; Hirabayashi 2015).  As rotational disruptions may potentially be as important in shaping the size distribution of main-belt asteroids as collisions (Jacobson et al.\ 2014), it will be crucial to understand as much as possible about the practical aspects of this process, where continued observational studies of rotationally disrupted asteroids will be very valuable in this regard.

\section{Future Outlook}

Despite the separate classifications of MBCs and disrupted asteroids, it should be emphasized that multiple mechanisms could contribute to any given dust emission event.  For example, an impact could trigger sublimation by excavating subsurface ice, or dust could be ejected by both weak sublimation and rapid rotation.  As such, as more objects are discovered, detailed studies to understand their unique characteristics will be essential before broader generalizations can be made.  Otherwise, perhaps the largest obstacle in the study of active asteroids is the extremely small number of known examples.  Surveys equipped with effective detection algorithms capable of finding these objects are thus vital to progress in this field (e.g., Waszczak et al.\ 2013; Hsieh et al.\ 2015).

\end{document}